\documentclass[12pt,leqno]{amsart}
 
\usepackage{amsmath}
\usepackage{graphicx}
\usepackage{color}
\input amssym.def
\input amssym

\long\def\symbolfootnote[#1]#2{\begingroup%
\def\thefootnote{\fnsymbol{footnote}}\footnote[#1]{#2}\endgroup}

\newtheorem{theorem}{\sc Theorem}[section] 
\newtheorem{lemma}[theorem]{\noindent {\sc Lemma}} 
\newtheorem{corollary}[theorem]{\sc Corollary}
\newtheorem{proposition}[theorem]{\sc Proposition}

\theoremstyle{plain}

\renewcommand{\a}{\alpha}
\renewcommand{\b}{\beta}
\renewcommand{\d}{\delta}

\newcommand{\e}{\varepsilon}
\renewcommand{\th}{\theta}
\newcommand{\g}{\gamma}
\newcommand{\G}{\Gamma}

\renewcommand{\L}{\Lambda}

\newcommand{\s}{\sigma}

\renewcommand{\t}{\tau}
\newcommand{\cal}{\mathcal}
\newcommand{\Z}{{\Bbb Z}}
\newcommand{\R}{{\Bbb R}}

\renewcommand{\o}{\omega}

\renewcommand{\i}{\infty}
\newcommand{\p}{\partial}

\renewcommand{\thefootnote}{\fnsymbol{footnote}}
\newcommand{\nat}{\natural}
\renewcommand{\thefootnote}{\fnsymbol{footnote}}

\begin{document}
 \author[J. Harrison  Department of Mathematics  U.C. Berkeley]{Jenny Harrison
\\Department of Mathematics
\\University of California, Berkeley}
\title[Hodge star]{Geometric Hodge Star Operator with Applications to the Theorems of Gauss and Green}
 
\begin{abstract}  The classical divergence theorem  for an $n$-dimensional domain $A$ and a smooth vector field $F$  in $n$-space  
$$\int_{\partial A} F \cdot n = \int_A div F$$   requires that a normal vector field $n(p)$ be defined a.e.  
$p \in \partial A$.   In this paper we give a new proof and extension of this theorem by replacing  $n$   with a 
limit $\star \partial A$ of $1$-dimensional polyhedral chains taken with respect to a norm.  The operator $\star$ is a geometric  dual to the Hodge star operator  and 
is defined on a large
class of $k$-dimensional domains of integration $A$  in $n$-space the author calls {\em chainlets}.   Chainlets include a broad range of domains, from smooth manifolds to  soap bubbles and
 fractals.     We prove as our main result the Star theorem
$$\begin{array}{rll}  
\int_{\star A} \omega &= (-1)^{k(n-k)}\int_A \star \omega. \end{array}$$ When combined with the general Stokes' theorem (\cite{Stokes}, \cite{continuity})
$$\int_{\partial A} \omega = \int_A d \omega$$ this result yields   
optimal and concise forms  of Gauss' divergence theorem
$$\int_{\star \partial A}\omega = (-1)^{(k-1)(n-k+1)} \int_A d\star \omega$$   and  Green's curl theorem
$$ \int_{\partial A} \omega =  \int_{\star A} \star d\omega.$$ 

\end{abstract}

\maketitle
 
\section{Introduction} 

In this paper we develop a theory of calculus on a large class of domains  by taking limits of $k$-dimensional polyhedral chains in $n$-space with respect to a one parameter family of norms depending on a parameter   $r \ge 0.$  
    Elements of the Banach spaces $\cal{N}_k^r$ obtained on completion are called   {\em $k$-chainlets} of class $N^r.$  The  norms  are decreasing with $r$.  The direct limit of the $\cal{N}_k^r$  is a normed linear space $\cal{N}_k^{\infty}.$  The  parameter $r$
reflects the {\em roughness} of the chainlets.  Concepts such as  smooth manifolds, fractals,
vector fields, differential  forms, foliations and measures have counterparts in chainlet geometry.    There is no geometric wedge product for all pairs of chainlets as this would lead to   multiplication of distributions.  However, other products and operators on differential forms do have dual geometric versions on chainlets  such as the Hodge star, Laplace and Dirac operators.

Chainlets of class $N^r$ are domains of integration for smooth
differential
$k$-forms $\o$ of class $B^r$, (i.e., the $r-1$ partial derivatives of $\o$ satisfy Lipschitz conditions).   The   geometric 
  Hodge star operator
 is a linear operator $\star $ from $k$-chainlets of class $N^r$ to $(n-k)$-chainlets of class $N^r$.
 It applies in all dimensions and codimensions and   does not require that tangents be defined anywhere.  Examples   include the subgraph
$A$ of the Weirstrass nowhere differentiable function $f$ defined over a compact interval, even though  $\p A$ has
infinite length   and has no tangents defined in the graph of $f$.    See Figure 1.  

\

\begin{figure}[b]\label{figure1}
 \begin{center}
 \resizebox{4.0in}{!}{\includegraphics*{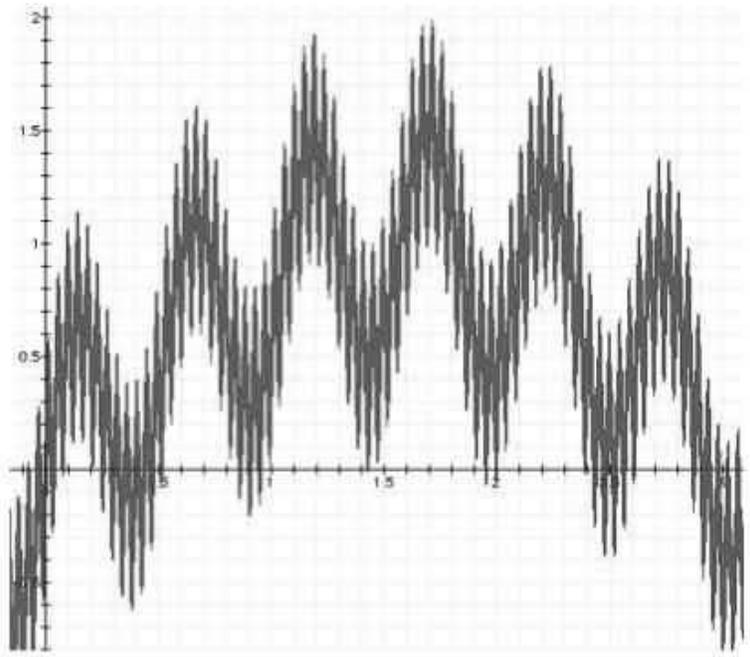}}
 \caption{The Weierstrass "nowhere differentiable" function}
\end{center}
 \end{figure}

\begin{theorem}[Star theorem]     If $A$ is a $k$-chainlet of class $N^r, r \ge 1,$  and $\o$  is  a differential $k$-form of class $B^r$  defined in a neighborhood of $spt(A)$    then   $$  \int_{\star A} \o = 
(-1)^{k(n-k)}\int_A \star \o.$$  
\end{theorem}

\begin{theorem}[General Stokes' theorem] \label{stokes}  If $A$ is a $k$-chainlet of class $N^r, r \ge 0,$ and $\o$  is a   differential $(k-1)$-form of class $B^{r+1}$   defined in a neighborhood of $spt(A)$  then  $$ \int_{\p A} \o  =  \int_A d \o.$$
\end{theorem}
This was first announced in \cite{Stokes} and proved in \cite{continuity}.  

Extensions of the
divergence and curl theorems for smooth differential forms and rough chainlets in
any dimension and codimension follow immediately.   
 
\begin{corollary}[General Gauss divergence theorem]\label{div}   If $A$ is a $k$-chainlet of class $N^r, r \ge 0$ and $\o$  is a   differential $(n-k+1)$-form of class $B^{r+1}$   defined in a neighborhood of $spt(A)$  then $$\int_{\star \p A} \o = (-1)^{(k-1)(n-k+1)} \int_A d\star \o.$$  
\end{corollary}

At one extreme, the form $\o$ must satisfy a Lipschitz condition (so that $d \star \o$ is bounded measurable) and be paired with a finite mass chainlet $A$  of class $N^0$, e.g., a polyhedral chain, for this theorem to be satisfied.   However, $\p A$ could have locally infinite mass as in Figure 1.    At the other extreme, if $\o$ is of class $C^{\i}$ the chainlet $A$ is permitted to have any degree of roughness, from soap films to fractals. 

Federer and de Giorgi \cite{federer}, \cite{gio} proved a divergence theorem for $n$-dimensional  currents C in $\R^n$
with $\cal{L}^n$ measurable support and a finite mass current boundary. The
vector field $F$ is assumed to be Lipschitz. $$\int F(x) \cdot n(C,x)d\cal{H}^{n-1} x = \int_C div F(x) d\cal{L}^n x.$$
The hypotheses imply the existence a.e. of measure theoretic normals
n(C, x) to the current boundary which is not required in 1.2.  Our result applies to all chainlets in the Banach spaces $\cal{N}_k^r$ which, in turn,  correspond to integrable currents
\cite{currents}.  These include all   currents satisfying  the hypotheses of the   theorem of Federer and de Giorgi.
  Federer wrote in the introduction to \cite{federer} \begin{quote} A striking application of our theory is the Gauss-Green type theorem ...
\end{quote} and  in the introduction to Chapter 4,  \begin{quote} Research on the problem of finding the most natural and general form of this theorem [Gauss-Green] has
contributed greatly to the development of geometric measure theory.
\end{quote}

\begin{corollary}[General Green's curl theorem]   If $A$ is a
$k$-chainlet of class
$N^r, r \ge 1,$ and $\o$ is  a  differential $(k-1)$-form of class $B^{r+1}$    defined in   $\R^n$  then $$    \int_{\p A} \o =   \int_{\star A} \star  d\o.$$  
\end{corollary}

A geometric coboundary operator $\diamondsuit$ for chainlets is defined as $$\diamondsuit := \star  \p \star $$ and a geometric Laplace operator $\Delta$ is defined
using combinations of   $\p$ and $\star$ :
$$\Delta A := (\p  + \diamondsuit)^2 A.$$    Let $\square$ denote the Laplace operator on  differential forms. 
\begin{corollary}[Laplace operator theorem] Let $r \ge1.$  If $A$ is a
$k$-chainlet of class
$N^r$ and $\o$  is  a  differential $k$-form  of class $B^{r+2}$    defined in a neighborhood of $spt(A)$   then $$  \int_{\Delta A} \o = (-1)^{n-1}\int_{A} \square \o.     $$  
\end{corollary}

 The norms are initially defined for polyhedral $k$-chains in Euclidean space $\R^n$ and it is shown at the end how to extend the results to singular $k$-chains in  Riemannian manifolds $M^n$.
 
 The main results in this paper were first announced in \cite{madeira}
\section{Norms on polyhedral chains}  
An {\em open half space} in $\R^k$ is the set of points which lie on a given side of a subspace of $\R^k$ of dimension $k-1$ and positively oriented.   A positively oriented open {\em $k$-cell} $\s$ of $\R^k$ is the nonempty, bounded intersection of a finite set of open half spaces $\R^k$.\symbolfootnote[2]{The standard approach assumes the half spaces are closed, but the open formulation is simpler and is sufficient for our purposes.}   If $\s_1$ and $\s_2$ are $k$-cells, so is their intersection $\s_1 \cap \s_2$, if it is nonempty.     
        In order to  define oriented $k$-cells in $\R^n$ we first orient the $k$-subspaces of $\R^n$ continuously.   An {\em oriented $k$-cell} $\s$ in $\R^n$ is an oriented $k$-cell in a $k$-dimensional subspace of $\R^n$.  A $0$-cell is a single point $\{x\}$ in $\R^n$.  (No
orientation need be assigned to the
 $0$-cell.)   Henceforth, all $k$-cells are assumed to be oriented.  The {\em support} $spt \s$ of a $k$-cell $\s$ is the set of all points  in the closure of the intersection of half spaces that determine $\s$.  

A {\em cellular $k$-chain}  $C$ is a  finite formal sum of $k$-cells $C = \sum_{i=1}^m a_i \s_i$ where the coefficients $a_i$ are taken to
be in
$G =
\Z$ or
$\R$.    We identify the oriented cell $\s$ with the chain $1 \s$.  If we let $-\s$ denote the cell $\s$ with the opposite orientation, we may set $-\s = (-1)\s$  and thus $a (-\s) = (-a) \s.$   

In order to define {\em polyhedral  $k$-chains} we form equivalence classes of the cellular $k$-chains.  
 Let $\s$ be an oriented cell.      Let $A$ be a cellular $k$-chain $A = \sum a_i \s_i$, written so that  the $\s_i$ are all positively oriented within their subspaces.   Define the function $A(x) := \sum a_i$ where the sum is taken over all $i$ such that $x \in spt \s_i.$   Set  $A(x) := 0$ if $x$ is not in the support of any $\s_i$.
 
We say that $k$ chains $A$ and $B$ are {\em equivalent} and write $A \sim B$ iff the functions $A(x)$ and $B(x)$ are equal except in a finite set of cells of dimension $< k.$   A {\em polyhedral  $k$-chain}  $P$ is an equivalence class so obtained\symbolfootnote[3]{ A
{\em polyhedral $k$-chain}
$P$ can also be defined as an equivalence classes of  
cellular $k$-chains
$C$ in 
$\R^n$ with   coefficients in $G$ where
$C_1
\sim C_2$ if and only if
$\int_{C_1 -C_2}
\o = 0$ for all smooth
$k$-forms $\o.$ }.  If $P$ is a chain $[P]$ denotes the polyhedral chain of $P$.  For example, $(-1,1) \sim (-1,0) + (0,1)$ and thus $ [(-1,1)] = [(-1,0) + (0,1)].$  In general,  if $\sum \t_{ij}$ is a subdivision of $\s_i$ then the polyhedral chains $\sum a_i \s_i$ and $\sum a_i \t_{ij}$ are equivalent.    As an abuse of notation we usually omit the square brackets and write $P$ instead of $[P]$.

   Note that every polyhedral chain $P$ has a nonoverlapping representative.
  
 \noindent{\bf Remark}. Two cells that have the same coefficient, but opposite orientation will cancel each other where they overlap.   If they have the same orientation, their coefficients are added.    
  
Denote the linear space of polyhedral chains by $\cal{P}_k.$   

The standard boundary operator $\p$ on $k$-cells $\s$ produces a $(k-1)$-chain.  This extends linearly to a boundary operator on cellular $k$-chains.  This, in turn, leads naturally to a well defined boundary operator $\p$ on polyhedral  $k$-chains  for
$k \ge 1$. For $k = 0 $ we set $\p P := 0$. Then
$$P_n \buildrel \p \over \to P_{n-1}
 \buildrel \p \over \to   \cdots   \buildrel \p \over \to P_1 \buildrel \p \over \to P_0$$
is a chain complex since $\p\circ \p = 0.$

\subsection*{Mass of polyhedral chains} \quad \\ Let $M(\s)$ denote k-dimensional Lebesgue measure, or k-volume of a $k$-cell
$\s$.   Every 0-cell $\s^0$ takes the form $\s^0 = \{x\}$ and we set $M(\s^0) = 1$.
  The {\em mass} of $P$ is defined by
$$M(P) := \sum_{i=1}^m
|a_i|M(\s_i)$$
where  $P = \sum_{i=1}^m a_i\s_i$ and the cells $\s_i$ are non-overlapping.  For example, the mass of a piecewise linear curve with multiplicity two is twice its arc length. Mass is a norm on the vector
space
$\cal{P}_k$.  Suppose $ \sum_{i=1}^m a_i \s_i$ is a non-overlapping representative of $P$.    The {\em support} of $P$
is defined as 
$spt(P) := \cup  
spt(\s_i).$   

It is worth noting to those well versed in analysis based on  unions and intersections of sets to note these definitions are substantially different and bring algebra of multiplicity and orientation into the mathematics at an early stage.

\subsection{The $k$-vector of a polyhedral chain} (\cite{whitney}, III)  The linear space of $k$-vectors in a vector space $V$ is denoted $\L^k(V).$   A $k$-vector is {\em simple} if it is of the form $v_1 \wedge \cdots \wedge v_k$.  A simple $k$-vector is a {\em $k$-direction} if it has unit mass.  A $k$-cell $\s$ determines a unique $k$-direction.  This, together with its $k$-volume $M(\s)$ determine a unique simple $k$-vector denoted $Vec(\s).$  Define the $k$-vector of a cellular $k$-chain $A = \sum a_i \s_i$ by $Vec(A) := \sum a_i Vec(\s_i).$    For $k = 0$ define $Vec(\sum a_i p_i): = \sum a_i.$   This definition extends to a polyhedral $k$-chain $P$ since the $k$-vector of any chain equivalent to a $k$-cell is the same as the $k$-vector of the $k$-cell.    

\begin{proposition}\label{zero}
If $P$ is a polyhedral $k$-chain then $Vec(\p P) = 0.$
\end{proposition}

\begin{proof}
This follows since $Vec(\p \s) = 0$ for every $k$-cell $\s$.
\end{proof}

\begin{theorem}\label{vec}  $Vec$ is a   linear operator
$$Vec: \cal{P}_k \to \L^k(\R^n)$$ with $$M(Vec(P)) \le M(P)$$ for all $P \in \cal{P}_k.$
\end{theorem}

\begin{proof}
This follows since $M(Vec(\s)) = M(\s)$ for every $k$-cell $\s$.
\end{proof}
  
  \subsection*{Natural norms}
                             
   For simplicity, we first define the norms in Euclidean space $\R^n$ and later show how to extend the definitions to Riemannian manifolds.  
 \subsection*{Diffchains}

For $v \in \R^n$ let $|v|$ denote its norm and $T_v$ translation through $v$.  Let   $\s^0$ be a $k$-cell in $\R^n$.  For consistency of terminology we also call $\s^0$ a {\em $(k,0)$-diffcell.} Let $v_1 \in \R^n$.  Define the {\em $(k,1)$-diffcell} $$\s^1 := \s^0 - T_{v_1}\s^0.$$  
This is a chain consisting of two cells, oppositely oriented.  A simple example consists of the sum of the opposite faces of a cube, oppositely oriented.  The chain is supported in these two faces.  Given $\s^0$ and $v_1, \cdots, v_r \in \R^n$, define the {\em $(k,j)$-diffcell} inductively $$\s^{j+1} := \s^j  - T_{v_{j+1}} \s^j.$$

 A {\em $(k, j)$-diffchain} $D^j$  in $\R^n$ is a finite sum of $(k, j)$-diffcells,
$$D^j =
\sum_{i=1}^m a_i \s_i^j$$
with coefficients $a_i \in G.$  (See Figure 2.) The vector space of all $(k, j)$-diffchains is
denoted  $\cal{D}_k^j.$

\begin{figure}[b]\label{figure2}
 \begin{center}
 \resizebox{4.0in}{!}{\includegraphics*{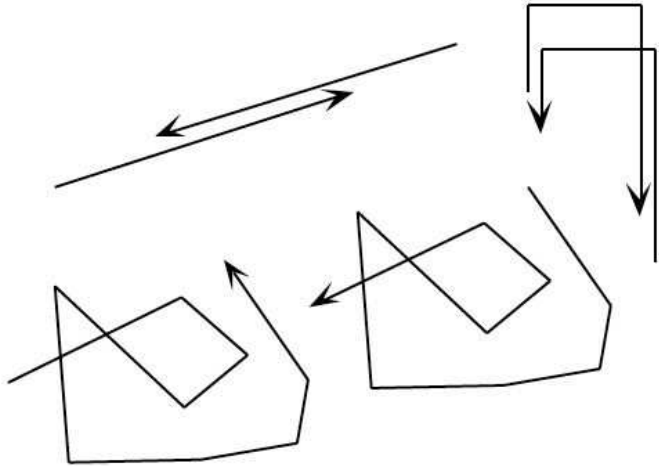}}
 \caption{A $(1,1)$-diffchain}
\end{center}
 \end{figure}

\subsection*{Diffchain mass} Given a $(k, j)$-diffcell $\s^j$ in $\R^n$ generated by a $k$-cell $\s^0$ and vectors $v_1,  \cdots , v_j$, define $\|\s^0\|_0 := M(\s^0)$ and for $j \ge 1$, 
$$\|\s^j\|_j := M(\s^0)|v_1| |v_2| \cdots |v_j |.$$
 For $D^j = \sum_{i=1}^m a_i \s_i^j$, possibly overlapping, define its {\em diffchain mass} as $$\|D^j\|_j := \sum_{i=1}^m |a_i|\| \s_i^j\|_j.$$

\subsection*{$r$-natural norms}
  Let $P \in \cal{P}_k$ be a polyhedral $k$-chain. For $r = 0$ define
$$|P|^{\natural_0} := M(P).$$
For $r \ge 1$ define the {\em r-natural} norm
$$|P|^{\natural_r} : = \inf\left\{\sum_{j=0}^r\|D^j\|_j + |C|^{\natural_{r-1}} \right\}$$  
where the infimum is taken over all decompositions
$$P = \sum_{j=0}^r D^j + \p C$$
where $D^j \in \cal{D}_k^j$ and $C  \in \cal{P}_{k+1}.$   It is clear $|\quad|^{\natural_r}$ is a semi-norm. 
 We shortly prove it is a norm.

It follows immediately from the definitions that the boundary operator
on chains is bounded w.r.t. the r-natural norms.
\begin{proposition}\label{lem.boundary} If $P \in  \cal{P}_k$ then
$$|\p P|^{\natural_{r}} \le  |P|^{\natural_{r-1}}.$$
\end{proposition}
\medskip \noindent {\bf Exercise} In the plane, define a sequence of polyhedral chains $P_k$ as follows:   Let $\s_k$ denote the positively oriented square centered at the origin with edge $2^{-k}.$  Let $P_k = 2^{2k} \s_k.$   Prove that the $P_k$ form a Cauchy sequence in the $1$-natural norm.  The boundaries $\p P_k$ form a Cauchy sequence in the $2$-natural norm.  
 
\section{Isomorphisms of differential forms and cochains}
We recall two classical results from integral calculus:
\begin{theorem}[Classical Stokes' theorem] \label{classicStokes} If $P$ is a polyhedral  $k$-chain and $\o$ is a smooth $k$-form defined in a neighborhood of $P$ then
$$\int_{\p P} \o = \int_P d\o.$$
\end{theorem}  

\begin{theorem}[Classical change of variables]\label{classicchange} If $P$ is a polyhedral  $k$-chain, $\o$ is a smooth $k$-form and $f$ is an orientation preserving diffeomorphism  defined in a neighborhood of $P$ then
$$\int_{f P} \o = \int_P f^* \o.$$

\end{theorem}

  \subsection*{The flat norm}
  Whitney's flat norm  on polyhedral chains $A \in \cal{P}_k$ is defined as follows:
$$|A|^{\flat}  := \inf\{M(B) + M(C): A = B + \p C,  B \in \cal{P}_k, C \in \cal{P}_{k+1}\}.$$

{\em Flat $k$-forms}  (\cite{whitney}, 12.4) are characterized as all bounded measurable $k$-forms $\o$ such that there exists a constant $C > 0$ with $\sup |\int_{\s} \o | < CM(\s)$ for all $k$-cells $\s$ and  $\sup |\int_{\p \t} \o | < CM(\t)$ for all $(k+1)$-cells $\t$.     The exterior derivative $d\o$ of a flat form $\o$ is defined a.e. and satisfies $$\int_{\p \t} \o = \int_{\t} d\o.$$

The {\em support} of a differential form is the closure of the set of all points $p \in \R^n$ such that $\o(p)$ is nonzero.   Let $U$ be an open subset of $\R^n$   and $\o$ be a bounded measurable $k$-form whose support is contained in    $U$.  In what follows, let $\s$ denote a $k$-cell and $\t$ a $(k+1)$-cell.
 
  Define $$\|\o\|_0 := \sup \left\{\frac{\int_{\s}\o}{M(\s)}: \s \subset spt \o \right\}.$$ 
  Inductively define $$\|\o\|_r := \sup\left\{\frac{\|\o -T_v\o\|_{r-1}}{|v|}:  spt(\o -T_v \o) \subset U\right\}.$$ 
  Define $$\|\o\|_0^{\prime} := \sup \left\{\frac{\int_{\p \t}\o}{M(\t)}: \t \subset spt \o\right\}$$  and   $$\|\o\|_r^{\prime} := \sup\left\{\frac{\|\o -T_v\o\|_{r-1}^{\prime}}{|v|}:  spt(\o -T_v \o) \subset U\right\}.$$

Define $$|\o|_0 := \|\o\|_0$$ and  for $r \ge 1$,
  $$|\o|_r := \max\{\|\o\|_o, \cdots, \|\o\|_r, \|\o\|_0^{\prime}, \cdots, \|\o\|_{r-1}^{\prime}\}.$$
   We say that $\o$ is of class $B^r$ if $|\o|_r < \i.$  Let $ \cal{B}_k^r$ denote the space of differential $k$-forms of class $B^r.$
  If $|\o|_1 < \i$ then $\o$ is a flat form.  It follows from    (\cite{whitney}, 12.4) that $d\o$ is defined a.e. and satisfies Stokes' theorem on cells: $$\int_{\p \s} \o = \int_{\s} d\o$$ yielding  
\begin{lemma} If $\o \in \cal{B}_k^r$, $r \ge 1$, then $$|\o|_r = \max\{\|\o\|_o, \cdots, \|\o\|_r, \|d\o\|_0 , \cdots, \|d\o\|_{r-1} \}.$$  
\end{lemma}

Therefore 
 
\begin{equation}\label{equation} |d\o|_{r-1} \le |\o|_r.
\end{equation}

The next result  generalizes the standard integral inequality of calculus:
$$\left|\int_P \o\right| \le M(P)|\o|_o$$ where $P$ is polyhedral and $\o$ is a bounded, measurable form.

\begin{theorem}[Fundamental integral inequality of chainlet geometry]\label{oldintegral} Let $P \in \cal{P}_k$, $r \in \Z^+,$ and $\o \in \cal{B}_k^r$ be defined in a neighborhood of $spt(P).$ Then
$$\left|\int_P \o\right| \le |P|^{\natural_r}|\o|_r.$$
\end{theorem}
 
\begin{proof}    We first prove $ \left|\int_{\s^{j}} \o \right| \le \|\s^j\|_{j}\|\o\|_{j}.$
Since $\|\o\|_0 = |\o|_0$ we know
$$\left|\int_{\s^0} \o \right| \le M(\s^0)|\o|_0 = \|\s^0\|_0\|\o\|_0. $$    
 
 Use the change of variables formula  \ref{classicchange} for the translation $T_{v_j}$ and induction   to deduce
$$\begin{array}{rll} \left|\int_{\s^{j}} \o \right| =  \left|\int_{\s^{j-1} - T_{v_j}\s^{j-1}} \o \right| 
&=  \left|\int_{\s^{j-1}} \o - T_{v_j}^* \o \right|  \\&\le  \|\s^{j-1}\|_{j-1}\|\o - T_{v_j}^*\o\|_{j-1} \\&\le  \|\s^{j-1}\|_{j-1}\|\o\|_{j}|v_{j}| \\&= \|\s^j\|_{j}\|\o\|_{j}\end{array}$$
 
  By linearity $$\left|\int_{D^j} \o\right| \le \|D^j\|_j \|\o\|_j$$
for all $(k, j)$-diffchains $D^j$.

  We again use induction to prove $\left|\int_P \o\right| \le  |P|^{\natural_r}|\o|_r.$ 
We know   $\left|\int_P \o\right| \le |P|^{\natural_0}|\o|_0.$  
Assume the estimate holds for $r-1.$ 

  Let $\e > 0$. There exists $P = \sum_{j=0}^r  D^j  + \p C$ such that $|P|^{\natural_r} >
\sum_{j=0}^r \|D^j\|_j + |C|^{\natural_{r-1}} - \e$. By Stokes' theorem for polyhedral chains, inequality (\ref{equation}) and
induction
 $$\begin{array}{rll}  |\int_P \o| &\le \sum_{j=0}^r |\int_{D^j} \o | + | \int_C d\o| \\& \le \sum_{j=0}^r \|D^j\|_j\|\o\|_j + |C|^{\natural_{r-1}} |d\o|_{r-1}\\& \le (\sum_{j=0}^r
\|D^j\|_j  + |C|^{\natural_{r-1}}) |\o|_r\\&\le  (|P|^{\natural_r} +
\e) |\o|_r.

\end{array} $$
Since the inequality holds for all $\e > 0$ the result follows.
\end{proof} 
\begin{corollary} $ |P|^{\natural_r}$ is a norm on the space of polyhedral chains $\cal{P}_k$. \end{corollary}
\begin{proof} Suppose $P \ne 0$ is a polyhedral chain. There exists a smooth differential form $\o$   such that $\int_P  \o \ne 0$. Then
$0 <  |\int_P \o| \le |P|^{\natural_r}|\o|_r$ implies $|P|^{\natural_r} > 0.$ \end{proof}
The Banach space of polyhedral    $k$-chains $\cal{P}_k$ completed with the
norm $|\quad |^{\natural_r}$ is denoted $\cal{N}^r_k $. The elements of $\cal{N}^r_k $ are called
{\em  $k$-chainlets of class} $N^r$.

It follows from Proposition \ref{lem.boundary} that the boundary $\p A$ of a k-chainlet $A$ of class $N^r$ is well defined as a $(k - 1)$-chainlet of class $N^{r+1}$.   If $P_i \to A$  in the $r$-natural norm define $$\p A:= \lim_{i \to \i} \p P_i.$$   By Theorem \ref{oldintegral} the integral $\int_A \o$ is well defined for k-chainlets $A$ of class $N^r$ and differential $k$-forms of class $B^r$.  If $P_i \to A$  in the $r$-natural norm define $$\int_A \o:= \lim_{i \to \i} \int_{P_i} \o.$$

\medskip \noindent{\bf Examples of chainlets}
\begin{enumerate}
\item {\em The boundary of any bounded, open subset $U$ of $\R^n$.}
One may easily verify that the boundary of the Van Koch snowflake supports a well defined chainlet of class $N^1$.  Suppose the frontier $\g$ of $U$ has positive Lebesgue area.   We are accustomed to finding the inner boundary and outer boundary of $U$ by approximating $g$ with polyhedral chains inside or outside $U$.    These curves are not identical as they bound a region with nonzero area.  We find two distinct chainlet boundaries of $U$ in this manner.   
 
 \item  {\em Graphs of functions} The graph  of a nonnegative $L^1$ function $f:K \subset \R^n \to \R$ supports a chainlet $\G_f$ if $K$ is compact.  This can be seen by approximating $\G_f$ by the polyhedral chains $P_k$ determined by a sequence of step functions $g_k$ approximating $f$.  The {\em subgraph} of a nonnegative function $f$ is the area between the graph of $f$ and its domain.  Since the subgraph of $f$ has finite area, the sequence $P_k$ is Cauchy in the natural norm.   The boundary $\p \G_f$ is a chainlet that identifies the discontinuity points of $f$.  
\end{enumerate}

\subsection*{Characterization of cochains as differential forms}

The $r$-natural norm of a cochain $X \in ({\cal N}^{r})^{\prime}$ is defined by
$$|X|^{\natural_r} := \sup_{P \in {\cal P}}\frac{|X \cdot P|}{|P|^{\natural_r}}.$$
The differential operator $d$  on cochains is defined as the dual to the
boundary operator $dX \cdot A := X \cdot \p A.$  This is the abstract version of Stoke's theorem.  The operator d is defined as the dual to the boundary operator and  Stokes' theorem becomes a definition in this category.   It remains to show how cochains relate to integration of differential forms and how the operator d given above relates to the standard exterior derivative of differential forms.   
  If
$X \in ({\cal N}_k^{r})^{\prime}$ then $dX \in ({\cal N}_{k+1}^{r-1})^{\prime}$
by Lemma \ref{lem.boundary}.

\subsection*{Cochains and differential forms}  In this section we show the operator $\Psi$ mapping differential forms of class $B^r$ into the dual space of chainlets of class $N^r$ via
integration
$$\Psi(\o)
\cdot A :=
\int_A \o$$ is a norm preserving isomorphism of graded algebras.      
 
It follows from   Theorem \ref{oldintegral} that$\Psi(\o) \in (\cal{N}_k^r)^{\prime}$ with
                            $$|\Psi(\o)|^{\natural_r} \le |\o|_{r}.$$

 \begin{theorem}[Extension of the theorem of de Rham] \label{theorem.iso}     Let $r \ge 0.$ To each cochain $X \in
\left({\cal N}_k^{r}\right)^{\prime}$ there corresponds a
unique differential form
$\phi(X)
\in {\cal B}_k^{r} $ such that $\int_{\s}\phi(X) = X\cdot \s$  for all
cells $\s$.
This correspondence is an isomorphism with
$$ |X|^{\natural_r} = |\phi(X)|_{r}.$$
 
 If $r \ge 1$ then
   $$\phi(dX) = d\phi(X).$$
 \end{theorem}

This is proved in \cite{iso}.   

\begin{corollary} \label{theorem.char} If $A, B \in {\cal N}_k^{r}$
satisfy $$\int_A \o =
\int_B
\o$$ for all $\o \in {\cal B}_k^{r}$ then $A = B$.
\end{corollary}

\begin{proof}  Let $X \in  ({\cal N}_k^{r})^{\prime}. $  By Theorem \ref{theorem.iso} the form $\phi(X)$ is of class $B^r$.  Hence
 $$X \cdot (A-B) = \int_{A-B}\phi(X) = 0.$$ It
follows that $A=B.$
\end{proof}

\begin{corollary} \label{theorem.norm} If $A \in   \cal{N}_k^r$ then

$$|A|^{\natural_r} = \sup\left\{\int_A \o :   \o \in B^r_k, |\o|_{r} \le 1\right\}.$$

\end{corollary}

\begin{proof}  By Theorem \ref{theorem.iso}

\[
\begin{array}{rll}
|A|^{\natural_r} &= \sup\left\{\frac{|X \cdot A|}{|X|^{\natural_r}}: X \in  (\cal{N}_k^r)^{\prime}\right\} \\&= 
\sup\left\{\frac{|\int_A \phi(X)|}{|\phi(X)|_{r}}: \phi(X)
\in   {\cal B}_k^{r}\right\} \\&=   \sup\left\{\frac{|\int_A
\o|}{|\o|_{r}}: \o\in   {\cal
B}_k^{r}\right\}.
\end{array}
\]
\end{proof}

\subsection*{Cup product} Given a $k$-cochain $X$ and a $j$-cochain $Y$, we
define their {\em cup product} as the $(j+k)$-cochain  
$$X\cup Y := \Psi(\phi(X) \wedge \phi(Y)).$$ It follows directly from Theorem
\ref{theorem.iso} that cup product corresponds to wedge product.

\begin{lemma}  Given  $X \in  (\cal{N}_k^r)^{\prime}$ and $Y \in  (\cal{N}_j^r)^{\prime}$  the cochain $X\cup Y \in  ({\cal
N}_{k+j}^{r}(R^n))^{\prime}$  with
                            $$|X\cup Y|^{\natural_r} = |\phi(X) \wedge \phi(Y)|_{r}.$$
Furthermore
                                $$\phi(X \cup Y) = \phi(X) \wedge \phi(Y).$$

\end{lemma}

\begin{theorem} If
  $X \in
({\cal N}_k^{r})^{\prime}, Y \in
({\cal N}_j^{r})^{\prime}, Z \in
({\cal N}_{\ell}^{r})^{\prime},$  and  $f
\in \cal{B}_0^{r+1}$ then
\begin{itemize}
\item[(i)] $|X \cup Y|^{\natural_r}
\le |X|^{\natural_r}|Y|^{\natural_r}$;
 \item[(ii)]    $d(X \cup Y) = dX \cup Y + (-1)^{j+k }  X \cup
dY;$
\item[(iii)]  $(X \cup Y) +
(Z \cup  Y) = (X+Z)\cup Y;$ and
\item[(iv)] $a(X \cup Y) = (aX \cup Y) = (X \cup aY).$

\end{itemize}
\end{theorem}

\begin{proof} These follow by using the isomorphism of
differential forms and cochains Theorem \ref{theorem.iso} and then applying corresponding results for
differential forms and their wedge products.

\end{proof}

Therefore the isomorphism $\Psi$ of Theorem \ref{theorem.iso} is one on graded
algebras. 

\subsection*{Continuity of {$ \mathbf {Vec(P)}$}}  

\begin{lemma}\label{Riemann}
Suppose $P$ is a polyhedral chain   and $\o$ is a bounded, measurable differential form.  If $\o(p) = \o_0$ for a fixed covector $\o_0$ and for all $p$ then $$\int_P \o = \o_0 \cdot Vec(P).$$
\end{lemma} 

\begin{proof}
This follows from the definition of the Riemann integral.
\end{proof}

  \begin{theorem}\label{massr}
If $P$ is a polyhedral $k$-chain then $$M(Vec(P)) \le |P|^{\nat_r} $$ for all $r \ge 1$
and 
$$ |P|^{\nat_1} \le M(Vec(P)) + \e M(P) \mbox{ if } spt(P) \subset B_{\e}(p) \mbox{ for some } p \in \R^n.$$    
\end{theorem}

\begin{proof}
Set $\a = Vec(P)$  and let $\eta_0$ be a covector such that $|\eta_0|_0 = 1$, and $\eta_0 \cdot \a = M(\a)$. Define the $k$-form $\eta$ by $\eta(p, \b) := \eta_0(\b).$  Since $\eta$ is constant it follows that $\|\eta\|_r = 0$ for all $r > 0$  and $\|d\eta\|_r = 0$ for all $r \ge 0.$  Hence $|\eta|_r = |\eta|_0  = |\eta_0|_0= 1.$
  By Lemma \ref{Riemann} and Theorem \ref{oldintegral} it follows that 
$$ M(Vec(P)) = \eta_0 \cdot Vec(P)  = \int_P \eta \le    |\eta|_r |P|^{\natural_{r}} = |P|^{\natural_{r}}.$$

For the second inequality we use Corollary \ref{theorem.norm}.  It suffices to show that $\frac{|\int_P \o|}{|\o|_1}$    is less than or equal the right hand side for any $1$-form $\o$ of class $B^1$.   Given such $\o$ define the $k$-form        $\o_0(q, \b) := \o(p,\b)$ for all $q$. By Lemma \ref{Riemann} 
\[
\begin{aligned}
 \left|\int_P \o\right| &\le \left|\int_P \o_0\right| + \left|\int_P \o -\o_0\right |  \\&\le
 |\o(p) \cdot Vec(P)| + \sup_q|\o(p) -\o(q)| M(P) \\& \le
   \|\o\|_0M( Vec(P))  +  \e \|\o\|_1 M(P) \\& \le
   |\o|_1( M(Vec(P)) + \e M(P))   \end{aligned}
\]
\end{proof}

If $A = \lim_{i \to \i} P_i$ in the $r$ natural norm then $\{P_i\}$ forms a Cauchy sequence in the $r$-natural norm.   By Theorem \ref{massr} $\{Vec(P_i)\}$ forms a Cauchy sequence in the mass norm on $\L^k(\R^n).$   Define $$Vec(A) := lim Vec(P_i).$$
\begin{corollary} \label{massrcor}  $$Vec: \cal{N}_k^r \to \L^k(\R^n)$$ is linear and continuous. 
\end{corollary}
   
\begin{corollary} \label{massrcor2}
Suppose $A$ is a chainlet of class $N^r$ and $\o$ is a differential form of class $B^r.$ If $\o(p) = \o_0$ for a fixed covector $\o_0$ and for all $p$  then $$\int_A \o = \o_0 \cdot Vec(A).$$ 
\end{corollary}

\begin{proof}  This is merely Lemma \ref{Riemann} if $A$ is a polyhedral chain.
   Theorem \ref{massr} lets us take limits in the $r$-natural norm.
\end{proof}

 \subsection*{The supports of a  cochain and of a chainlet}   The
support $spt( X)$ of a cochain $X$ is the set of points $p$ such that for each $\e > 0$ there is a cell $\s  \subset
U_{\e}(p)$ such that $X\cdot \s \ne 0.$

 The support $spt(A)$ of a chainlet $A$ of class $N^r$  is the set of points $p$ such that for each $\e > 0$ there is a cochain $X$ of class $N^r$ such that $X \cdot A \ne 0$ and $X \cdot \s = 0 $ for each $\s$  supported outside $U_{\e}(p).$   We prove that this coincides with the definition of  the support  of $A$ if $A$ is a polyhedral chain. Assume $A = \sum_{i=1}^m a_i \s_i$ is nonoverlapping and the $a_i$ are nonzero.  We must show that $spt(A)$ is the union $F$ of the $spt(\s_i)$ using this new definition.    Since $X \cdot A = \int_{A}  \phi(X)$  it follows that $spt(A) \subset F.$   Now suppose $x \in F$; say $x \in \s_i.$  Let $\e > 0.$  We find easily a smooth differential form $\o$ supported in $U_{\e}(p)$, $\int_{\s_i} \o \ne 0$, $\int_{\s_j} \o = 0, j \ne i$.  Let $X$ be the cochain determined by $\o$ via integration.   Then $X \cdot A \ne 0$     and $X \cdot \s = 0 $ for each $\s$  supported outside $U_{\e}(p).$  

\begin{proposition}\label{spt}  If $A$ is a chainlet  of class $N^r$ with  $spt(A)= \emptyset$ then $A = 0.$  If $X$ is a cochain of class $N^r$ with   $spt(X) = \emptyset $ then $ X = 0.$
\end{proposition}

\begin{proof}  By Corollary \ref{theorem.norm} suffices to show $X \cdot A = 0$ for any cochain $X$ of class $N^r$.  Each $p \in spt(X)$ is in some neighborhood $U(p)$ such that $Y \cdot A = 0$ for any $Y$ of class $N^r$ with $\phi(Y) = 0$ outside $U(p)$.    Choose a locally finite covering  $\{U_i, i \ge 1\}$  of $spt(X)$.  Using a partition of unity $\{\eta_i\} $  subordinate to this covering we have $$X = \sum \eta_i X$$ and $\phi(\eta_i X) = \eta_i \phi(X) = 0$ outside $U_i$.  Hence $$X \cdot A = \sum (\eta_iX \cdot A) = 0.$$

For the second part it suffices to show that $X \cdot \s = 0$ for all simplexes $\s$.  Each $p \in \s$ is in some neighborhood $U(p)$ such that $X \cdot \t = 0$ for all $\t \subset U(p).$  We may find a subdivision $\sum \s_i$ of $\s$ such that each $\s_i$ is in some $U(p)$.  Therefore $X \cdot \s = \sum X \cdot \s_i = 0.$
\end{proof}  

\section{Geometric star operator}

\subsection*{Differential $k$-elements} In this section we make precise the notion of an {\em infinitesimal} of calculus.   Imagine taking an infinitely thin square card and cutting it into four pieces.  Stack the pieces and repeat, taking a limit.  What mathematical object do we obtain?  The reader will recall monopoles and Dirac delta functions which are closely related.  We show the limit, the author calls a {\em  differential $k$-element}, exists as  a well defined chainlet and thus may be acted upon by any chainlet operator.      We emphasize that these operators are geometrically defined as opposed to the duals of operators on differential forms.

 Let $p \in \R^n$  and $\a$ be a  $k$-direction in $\R^n$.    A  {\em unit  differential $k$-element} $\a_p$ is defined as follows:
For each $\ell \ge 0$, let $Q_{\ell} = Q_{\ell}(p,\a)$ be the weighted $k$-cube centered at $p$ with $k$-direction $\a$, edge $2^{-\ell}$ and  coefficient $2^{k \ell}$.  
Then
$M(Q_{\ell}) =1.$ 
  We show that $\{Q_{\ell}\}$ forms a Cauchy sequence in the $1$-natural norm.    Let $j \ge 1$ and estimate $|Q_{\ell} -Q_{\ell + j}|^{\natural_1}.$  Subdivide
$Q_{\ell}$ into
$2^{kj}$ binary cubes $Q_{\ell,i}$ and consider $Q_{\ell +j}$ as $2^{kj}$ copies of   $\frac{1}{2^{kj}}Q_{\ell +j}.$   We form $(k, 2)$-diffcells  (bicells) of these subcubes of
$Q_{\ell} -Q_{\ell + j}$ with translation distance $\le 2^{-\ell}.$   Since the  mass of each
$Q_{\ell}$ is one, it follows that 
$$| Q_{\ell} -Q_{\ell + j}|^{\natural_1}  = \left|
\sum_{i = 1}^{2^{kj}} \left(Q_{\ell,i} - \frac{1}{2^{kj}}Q_{\ell + j}\right)\right|^{\natural_1}
\le 
  2^{-\ell}.$$
 Thus $Q_{\ell}$ converges to a $1$-natural chain denoted $\a_p$  with $|\a_p - Q_{\ell}|^{\natural_1} \le 2^{1-\ell }.$   If we let $\a$ be any $k$-vector with nonzero mass, the same process will produce a chainlet $\a_p$ whose mass is the same as that of $\a$ and depends only on $\a$ and $p$.    If $\o$ is a Lipschitz form defined in a neighborhood of $p$ then $\int_{\a_p} \o = \o(p; \a) $ by   Theorem \ref{oldintegral}

The next lemma shows the definition of a differential $k$-element is well defined. 
 \begin{proposition}\label{vecprop}  Fix a  nonzero $k$-vector $\a$ and $p \in \R^n.$  Let $\{P_i\}$ be a sequence of polyhedral $k$-chains such that $$M(P_i) \le C,   spt(P_i) \subset B_{\e_i}(p), Vec(P_i) \to \a$$ for some $C > 0$ and $\e_i \to 0$.
 Then there exists a unique nonzero chainlet $$A = \lim_{i \to \i} P_i$$ in the $1$-natural norm with $Vec(A) = \a$ and $spt(A) = \{p\}.$
\end{proposition}

\begin{proof} By Theorem \ref{massr} 
\[
\begin{aligned} |P_i - P_j |^{\natural_{1}} &\le M(Vec(P_i) -  Vec(P_j)) + \e M(P_i - P_j)
   \to 0.
\end{aligned}
\]
By Corollary \ref{massrcor} there exists $A = \lim_{i \to \i} P_i$ in the $1$-natural norm with $Vec(A) = \a.$  Therefore $A \ne 0.$  (Else, $Vec(A) = 0.$)

By the definition of support,  $spt(A)$ is either the empty set or the set $\{p\}.$  By Proposition \ref{spt} $spt(A) = \emptyset \implies A = 0.$   
\end{proof}

\begin{theorem} Fix $p\in \R^n.$ The operator $$Vec: \cal{N}_k^r \to \L^k(\R^n)$$ is one-one on chainlets supported in   $p$.
 \end{theorem}

\begin{proof}    By Proposition \ref{vecprop} and 
Theorem \ref{vec}   we only need to show that if $Vec(A) = 0$ then $A = 0.$  Let $X$ be an $r$-natural cochain.  Define $X_0$ by $$\phi(X_0)(q) := \phi(X)(p)  \mbox{ for all } q.$$   By Corollary \ref{massrcor2} 
$$X \cdot A = X_0 \cdot A = \phi(X)(p) \cdot Vec(A) = 0$$ implying $A = 0.$ 
\end{proof}

In \cite{discrete} and \cite{ravello} we will develop the discrete theory more fully.  The boundary of a differential $k$-element will be studied, as well as actions of other operators.  The full calculus may be developed starting with differential $k$-elements   replacing  $k$-dimensional tangent spaces with differential $k$-elements. 

An {\em elementary $k$-chain} $\dot{P} = \sum_{i=1}^m b_i (\a_p)_i$   is a chain of   differential $k$-elements $ (\a_p)_i$ with coefficients $b_i$ in $G$. (Note that both the $k$-vector $\a$ and point $p$ may vary with $i$.)   Denote the vector space of elementary $k$-chains in
$\R^n$ by $\cal{E}_k.$   

\begin{theorem}[Density of elementary chains] \label{thmdiscrete} The space of  elementary $k$-chains $\cal{E}_k$ is dense in $\cal{N}_k^r.$
\end{theorem}

\begin{proof}   Let $R$ be a unit $k$-cube in $\R^n$ centered at $p$ with $k$-direction $\a.$   For each $j \ge 1 $  subdivide $R$
into $2^{kj}$ binary cubes  $R_{j,i}$ with midpoint $p_{j,i}$  and edge $2^{-j}.$  Since $R_{j,i} = 2^{-jk} Q_j(p_{j,i},\a)$  it follows that
$$
\begin{array}{rll} |R_{j,i} -   2^{-jk}  \a_{p_{j,i}}|^{\natural_1} &=   2^{-jk}|Q_j(p_{j,i},\a) -    \a_{p_{j,i}}|^{\natural_1} \\&\le  
 2^{-jk} 2^{-j+1} =  2^{-j+1}M(R_{j,i}).\end{array}$$  Let $\dot{P}_j = \sum_{i=1}^m 2^{-jk}\a_{p_{j,i}}.$    Then 
$$|R -  \dot{P}_j|^{\natural_1} \le    2^{-j+1}\sum M(R_{j,i}) = 2^{-j+1}M(R) = 2^{-j+1}.$$

This demonstrates  that  $\dot{P}_j  \buildrel \natural_{1} \over \to R$.  This readily extends to any cube with edge $\e$. 

Use the Whitney decomposition to subdivide a $k$-cell $\t$ into binary $k$-cubes.   For each $j \ge 1$ consider the finite sum of these cubes    with edge $\ge 2^{-j}.$    Subdivide each of these cubes into subcubes $Q_{ji}$ with edge $2^{-j}$  obtaining  $\sum_i Q_{ji} \to \t $ in the mass norm as $j \to \i$.      Let $\a = Vec(\t)$ and $p_{ji}$ the midpoint of $Q_{ji}.$    Then 
$$|\t - \sum_i \a_{p_{ji}}|^{\natural_{1}} \le |\t - \sum_i  Q_{ji}|^{\natural_{1}}  +
\sum_i| Q_{ji} -  \a_{p_{ji}}|^{\natural_{1}}.$$    We have seen that the first term of the right hand side tends to zero as $j \to \i.$  The second is bounded by $\sum_i M(Q_{ji}) 2^{-j+1} < M(\t) 2^{-j+1} \to 0.$    It follows that 
$\t$ is approximated by elementary $k$-chains   in the $1$-natural norm. 
   Thus elementary $k$-chains are dense in $\cal{P}_k.$  The result follows since polyhedral chains
are dense in chainlets.

\end{proof}

\subsection*{Geometric Hodge star}
We next define a geometric Hodge star operator on chainlets.     If $\a$ is a $k$-direction in $\R^n$ then $\star \a$ is defined to be the $(n-k)$-direction orthogonal to $\a$ with complementary 
orientation. Define $$\star( \a_p) := (\star \a)_p.$$   We may omit the parentheses without ambiguity.  The operator $\star$ extends to   elementary $k$-chains $\dot{P}$ by linearity.   It follows immediately that $\int_{ \dot{P}} \o =
\int_{\star \dot{P}} \star \o$.  By Theorem \ref{theorem.norm}
$|\star \dot{P}|^{\natural_r} = |\dot{P}|^{\natural_r}.$    We may therefore define $\star A$ for any chainlet $A$ of class $N^r$ as follows:  By Theorem \ref{thmdiscrete} there exists elementary $k$-chains $\{\dot{P}_j\}$ such that $A = \lim_{j \to \i} \dot{P}_j$ in the $r$-natural norm.   Since  $\{\dot{P}_j\}$ forms a Cauchy sequence we know $\{\star \dot{P}_j\}$ also forms a Cauchy sequence.  Its limit in the $r$-natural norm is denoted $\star A.$    This definition is independent of the choice of the sequence $\{\dot{P}_j\}.$  (See Figure 3 for an example.)
       
\begin{figure}[b]\label{figure3}
 \begin{center}
 \resizebox{4.0in}{!}{\includegraphics*{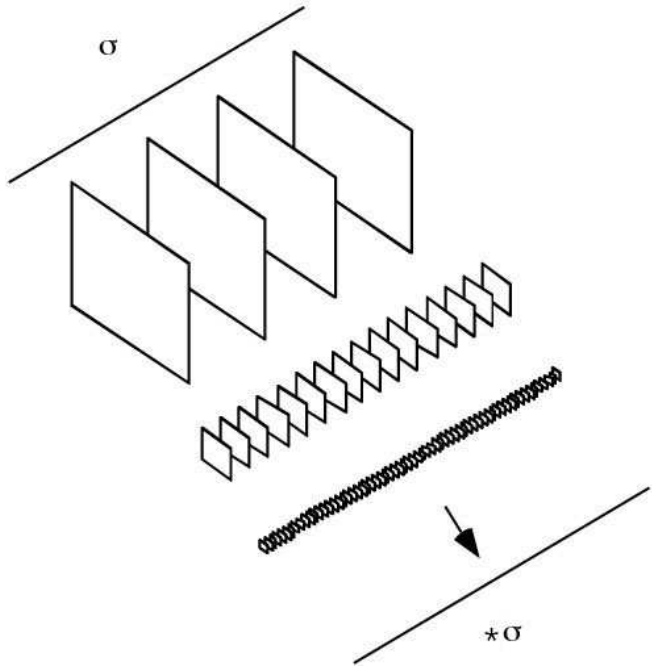}}
 \caption{Hodge star of a 1-simplex in 3-space}
\end{center}
 \end{figure} 
 
   \begin{theorem}[Star theorem]   \label{thm.star}
$\star  : {\cal N}_k^{r} \to{\cal N}_{n-k}^{r} $ is a
 norm-preserving linear  
operator that is adjoint to the Hodge star operator on forms.  It  satisfies $\star  \star  =  (-1)^{k(n-k)} I$ and 
 
 $$\int_{\star  A}
\o = (-1)^{k(n-k)} \int_A \star \o$$   for all $A \in \cal{N}_k^r$ and
 all $(n-k)$-forms $\o$ of class $B^r$ defined in a neighborhood of $spt(A).$
\end{theorem}
   
   \begin{proof}  We first prove this for differential $k$-elements $\a_p.$   Since  $\a_p$ is a $1$-natural chainlet we may integrate $\o$ over it.  Hence $$\int_{\a_p} \o = \o(p;\a) = \star \o(p; \star \a) = \int_{\star \a_p} \star \o.$$ It follows that $ \int_{\dot{P}}  \o = \int_{\star \dot{P}} \star \o $ for any elementary $k$-chain $\dot{P}$.  Let $A$ be a chainlet of class $N^r$. It follows from Theorem \ref{thmdiscrete} that $A$ is approximated by elementary $k$-chains $ A = \lim_{j \to \i} \dot{P}_j$ in the $r$-natural norm.  We may apply continuity of the integral (Theorem \ref{oldintegral}) to deduce 
   $$\int_{A} \o = \int_{\star A} \star \o.$$

The Hodge star operator on   forms satisfies $\star \star  \o = (-1)^{k(n-k)} \o.$  It follows that $$\int_{A} \star \o  = \int_{\star A} \star \star  \o = (-1)^{k(n-k)}\int_{\star A} \o.$$

   \end{proof} 
     
 \subsection*{Geometric coboundary of a chainlet}
Define the {\em geometric  coboundary } operator $$\diamondsuit: \cal{N}_k^r\to \cal{N}_{k+1}^{r+1}$$  by $$\diamondsuit  : =
\star  \p
\star.$$ 
 \begin{figure}[b]\label{figure4}
 \begin{center}
 \resizebox{4.0in}{!}{\includegraphics*{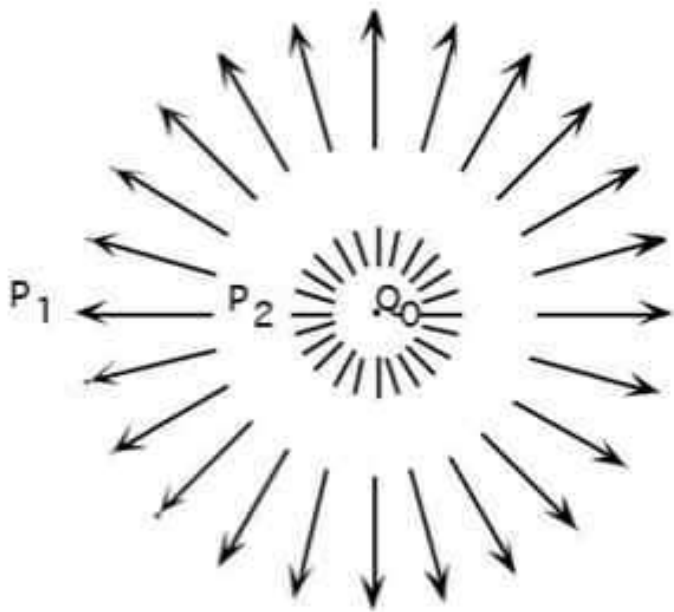}}
 \caption{Geometric coboundary of a  point $Q_0$  as a limit of polyhedra $P_k$}
\end{center}
 \end{figure} 

The following theorem follows immediately from properties of boundary $\p$ and star $\star $.  Let $\d:= \star d \star$ denote the coboundary  operator on differential forms.  

\begin{theorem}[Coboundary theorem] $\diamondsuit : \cal{N}_k^r \to {\cal N}_{k+1}^{r+1}$ is a nilpotent linear operator
satisfying
 \begin{itemize}
\item[(i)] $\int_{\diamondsuit A} \o = (-1)^{n-1} \int_A  \delta \o$  for all $\o$ defined in a neighborhood of $spt(A)$;
\item[(ii)] $\star  \p A = (-1)^{k(n-k)}\diamondsuit\star  A;$ and
\item[(iii)] $|\diamondsuit A|^{\natural_r} \le |A|^{\natural_{r-1}}$ for all chainlets $A$. 
\end{itemize} 
 \end{theorem}

\subsubsection*{Geometric interpretation of the coboundary of a chainlet} This has a geometric
interpretation seen by taking  approximations by polyhedral chains.   For example,  the
coboundary
of $0$-chain $Q_0$ in $\R^2$ with  unit $0$-mass and supported in
a single point $\{p\}$  is the limit of $1$-chains $P_k$ depicted in Figure 4.

The coboundary of a $1$-dimensional unit cell $Q_1$   in $\R^3$  is
approximated by a ``paddle wheel'',
 supported in a neighborhood of  $|\s|$.

If  $Q_2$ is a unit $2$-dimensional square in
$\R^3$ then its coboundary $\diamondsuit Q_2$ is approximated by the sum of
two weighted sums of oppositely oriented pairs of small $3$-dimensional
balls, 
one collection slightly above
$Q_2$, like a mist, the other collection slightly below $Q_2.$     A snake approaching the
boundary of a lake knows when it has arrived.  A bird
approaching the coboundary of a lake knows when it has
arrived.
 
\subsection*{Geometric Laplace operator}

The geometric Laplace operator $$\Delta: {\cal N}_k^{r}
\to {\cal N}_k^{r+2} $$ is defined on chainlets
by $$\Delta A := (\p + \diamondsuit)^2 A = (\p \diamondsuit + \diamondsuit \p)A.$$

\begin{theorem}[Laplace operator theorem] Suppose $A \in {\cal N}_k^{r}$ and  $\o \in {\cal
B}_k^{r+2} $ is defined in a neighborhood of $spt(A).$ Then $\Delta A \in  {\cal N}_k^{r+2}$,  $$|\Delta A|^{\natural{r+2}} \le |A|^{\natural_r},$$ and $$\int_{\Delta A} \o = (-1)^{n-1}\int_{A} \Delta \o.$$ 
\end{theorem}

 The geometric Laplace operator on chainlets requires at least the
$2$-natural norm.   Multiple iterations of
$\Delta$ require   the $r$-natural norm for larger and larger $r$.
For spectral analysis and applications to dynamical systems the normed linear space
${\cal N}_{k}^{{\i}}$  with the operator  $$\Delta: {\cal N}_{k}^{{\i}} \to {\cal
N}_{k}^{{\i}}$$ should prove useful.  (See \cite{ravello} for further discussion of the direct limit space ${\cal N}_{k}^{{\i}}$.)

A chainlet is   {\em harmonic} if   $$\Delta A = 0.$$  It should be of considerable interest to 
study the spectrum of the   geometric Laplace operator $\Delta$ on chainlets.\symbolfootnote[1]{The
geometric Laplace operator was originally defined by the author with the object of developing a geometric Hodge theory.  A nice step of this project can be found in J. Mitchell's
Berkeley thesis, drafted under the author's supervision
\cite{Mitchell}.}     
 
 \subsection*{Geometric representation of
differentiation of distributions}
 
 An {\em $r$-distribution} on $\R^1$ is a bounded
linear functional on functions $f \in {\cal B}_0^{r}(\R^1)$  with compact support. 
Given a one-dimensional  chainlet $A$ of class $N^r$ 
define an  $r$-distribution
$\th = \th (A)$  by  $\th(f) := \int_A f (x)dx$,   for
$f \in {\cal B}_0^{r}(\R^1).$

\begin{theorem}The mapping $\th$ is an
endomorphism. Furthermore, differentiation in the sense of distributions
corresponds geometrically to the operator $ \star \p $.   That is, if $\th(A) =
\g_A$ then $\th(\star \p A) = \g_A'.$\symbolfootnote[3]{Since this paper was first submitted, the author has extended this result to currents. \cite{currents}} 
\end{theorem}

\begin{proof}
Suppose $\th(A) = \th(B)$.  Then $ \int_A
f(x)dx = \int_B f(x)dx$ for all functions $f \in
{\cal B}_0^{r}.$  But all $1$-forms $\o \in {\cal
B}_1^{r}$ can be written $\o = f dx.$  By Corollary \ref{theorem.char}
chainlets are determined by their integrals and    thus  
   $A = B$.
 
  We next show that $\g_{\star \p A} =
\g_A^{\prime}.$  Note that $\star (f(x) dx) = f(x).$  Thus
\[
\begin{array}{rll}
\g_{\star \p A}(f) &= \int_{\star \p A} f(x)dx  = \int_{\p A}
f = \int_A df \\&= \int_A f^{\prime}(x)dx =
\g_A(f^{\prime}) = \g_A^{\prime}(f).
\end{array}
\]
\end{proof}

\section{Extensions of   theorems of Green and Gauss}

\subsection*{Curl of a vector field over a chainlet}

Let $S$ denote a smooth, oriented surface with boundary in
$\R^3$ and $F$ a smooth vector field defined in a neighborhood of $S$.  The usual way to integrate the curl of a vector
field $F$ over $S$ is to integrate the Euclidean dot product of
curl$F$ with the unit normal vector field of $S$ obtaining    $\int_S curlF \cdot n dA$.  By the curl theorem this integral equals
$\int_{\partial S} F \cdot  d\s.$ 

We translate this into the language of chainlets and differential forms.

Let $\o$ be the unique differential $1$-form associated to $F$
by way of the Euclidean dot product.   The differential form version of $curl F $ is    $\star d\o.$  
The  unit normal vector field of $S$ 
 can be represented as the chainlet $\star S$.  Thus the net curl of $F$ over
$S$ takes the form  $\int_{\star S} \star  d\o.$ By the Star theorem \ref{thm.star} and Stokes' theorem for chainlets \ref{stokes}  this integral equals $\int_S d\o = \int_{\p S}
\o.$ The vector version of the right hand integral is  $\int_{\partial S} F \cdot  ds. $  The following extension of Green's curl theorem to chainlets of arbitrary dimension and codimension follows immediately from Stokes' theorem and the Star theorem and is probably optimal.

\begin{theorem}[General Green's curl theorem]  Let $A$ be a  $k$-chainlet of class $N^r$ and $\o$ a differential $(k-1)$-form of class $B^r$ defined in a neighborhood of $spt(A).$  Then $$\int_{\star A} \star  d \o =
\int_{\p A}
\o.$$ 
\end{theorem}  
 
\begin{proof}
This is a direct consequence of  Theorems \ref{stokes} and \ref{thm.star}.
\end{proof}
It is not necessary for  tangent spaces  to exist for $A$ or $\p A$ for this theorem to hold.

\subsection*{Divergence of a vector field over a chainlet} \medskip The usual way to calculate divergence of a vector field
$F$ across a boundary of a smooth surface
$D$ in $\R^2$  is to integrate the dot product of $F$ with
the unit normal vector field of
$\partial D$.   According to
Green's  Theorem, this quantity equals  the
integral of the divergence of $F$ over $D$.  That is, 
$$\int_{\partial D} F \cdot nd\s = \int_D divF dA.$$   Translating this into the language of differential forms and chainlets with an appropriate sign adjustment, we
  replace the unit normal vector field over $\p D$ with the chainlet $\star  \p D$ and $div F$ with the differential form $d \star \o.$ 
 We next give an extension of the Divergence theorem to $k$-chainlets in $n$-space.  As before, this follows immediately from Stokes' theorem and the Star theorem and is probably optimal.

\begin{theorem}[General Gauss divergence theorem]  
Let $A$ be a  $k$-chainlet of class $N^r$   and $\o$ a differential $(n-k+1)$-form of class $B^{r+1}$    defined in a neighborhood of $spt(A)$
then

 $$\int_{\star  \partial A} \o = (-1)^{(k-1)(n-k-1)} \int_{A} d\star \o.$$
\end{theorem}

\begin{proof}
This is a direct consequence of  Theorems \ref{stokes} and \ref{thm.star}.
\end{proof}

As before, tangent vectors need not be defined for the theorem to be valid and it holds in all dimensions and codimensions.

\subsection*{Riemannian manifolds}   In order to extend the definitions to smooth Riemmanian manifolds    replace cells $\s^0$ with singular cells  $\t^0 = f\s^0$.   Define $M(\t^0)$  to be $k$-dimensional Hausdorff measure of $\t^0$.      Replace vectors $v$ with smooth, divergence free vector fields $v$ defined in a neighborhood of $spt \t$ and let  $T_v$ denote the time one map of the flow of $v$.  Define $$|v| := \sup\{|v(p)|: p \in spt v\}.$$   Norms of differential forms are defined as before, replacing vectors with vector fields.   The previous definitions and results carry through locally.  Global results require the pushforward operator be defined for chainlets and a change of variables result.  These are   naturally established using differential $k$-elements and deduced for chainlets by taking limits.   

If 
$f:U\subset \R^n \to V \subset \R^n$ is Lipschitz   
then $Df_p$ is defined on a subset of full measure   by Rademacher's theorem.  Therefore, for a.e. $p \in U,$  
$$f_*(\a_p) :=  Df^k_p(\a_p)$$    is well defined so that $f^* \o(p, \a_p) = \o(f(p), f_* \a_p).$   Passing to integrals, we have $$\int_{f_* \a_p} \o = \int_{\a_p} f^*\o.$$  Therefore, 

\begin{proposition}\label{change} $$\int_{f_* \dot{P}} \o = \int_{\dot{P}} f^*\o$$ for all elementary chains $\dot{P}.$   
\end{proposition}

\begin{proposition}  Let $f: U \subset \R^n \to \R^n$ be a mapping of class $B^{r+1}$ and $\dot{P}$ be an elementary chain.  Then  $$|f_*\dot{P}|^{\natural_{r}} \le |f|_{r+1}|\dot{P}|^{\natural_{r}}.$$  
\end{proposition}

\begin{proof}  By Theorem \ref{oldintegral} $$\left|\int_{f_*\dot{P}} \o \right|  = \left| \int_{\dot{P}} f^* \o\right| \le |f^*\o|_r |\dot{P}|^{\natural_{r}} \le  |f|_{r+1}|\o|_r|\dot{P}|^{\natural_{r}}.$$
\end{proof}

Since elementary chains are dense in chainlets we may define $$f_*A := \lim_{i \to \i} f_* \dot{P_i}$$ where $A = \lim_{i \to \i} \dot{P_i}$ in the $r$-natural norm.   We deduce
\begin{theorem}[Pushforward operator]
Let $f: U \subset \R^n \to \R^n$ be a mapping of class $B^{r+1}$ and $A$ be a chainlet of class $N^r$.  Then  $$|f_*A|^{\natural_{r}} \le |f|_{r+1}A|^{\natural_{r}}.$$ 
\end{theorem}
We close with a concise and general change of variables formula.  
\begin{theorem}[Change of variables] Let $f: U \subset \R^n \to \R^n$ be a mapping of class $B^{r+1}$, $\o$ be a differential form of class $B^r$ and $A$ be a chainlet of class $N^r$. Then $$\int_{f_*A} \o = \int_A f^* \o.$$
\end{theorem}

This follows by Proposition \ref{change} and taking limits in the chainlet norm.

\end{document}